\documentstyle[amsfonts,aps]{revtex}

\begin{document}
\title{Functional Approach to Quantum Decoherence and the Classical Final Limit:
the Mott and Cosmological problems.}
\author{Mario Castagnino}
\address{Instituto de Artronom\'{\i}a y F\'{\i}sica del Espacio\\
Casilla de Correos 67, Sucursal 28\\
1428, Buenos Aires , Argentina.}
\author{Roberto Laura}
\address{Instituto de F\'{\i}sica de Rosario.\\
Av. Pellegrini 250,\\
2000 Rosario, Argentina.}
\maketitle

\begin{abstract}
Decoherence and the approach to the classical final limit are studied in two
similar cases: the Mott and the Cosmological problems.
\end{abstract}

\section{Introduction.}

One of the most important problems of theoretical physics in the last years
was to answer the question: How and in what circumstances a quantum system
becomes classical ? \cite{WyZ}. In spite of the great effort made by the
physicists to find the answer, the problem is still alive \cite{Giuliani}
and we are far from a complete understanding of many of its most fundamental
features. In fact the most developed and sophisticated theory on the
subject, histories decoherence, is not free of strong criticisms \cite
{Dowker}.

For conceptual reason we will decompose the limit {\it quantum mechanics }$%
\rightarrow $ {\it classical mechanics }in two processes: {\it quantum
mechanics }$\rightarrow $ {\it classical statistical mechanics }and {\it %
quantum statistical mechanics }$\rightarrow $ {\it classical mechanics. }%
There is an almost unanimous opinion that the first process is produced by
two phenomena:

i. {\it Decoherence,} that in quantum systems, restores the boolean
statistic typical of quantum mechanics and

ii.-{\it The limit }$\hbar \rightarrow 0,${\it \ }that circumvents the
uncertainty relation at the macroscopic level and allows to find the
classical behaved density functions via de Wigner integral.

The second process is produced by these phenomena plus the localization (or
production of correlation) phenomenon \cite{LauraF}. We will discuss at
large the first process and briefly deal with the second at the end of each
section.

The techniques to deal with the two phenomena related with the first process
are not yet completely developed. One of the main problems is to find a
proper and unambiguous definition of the, so called, {\it pointer basis,}
where, decoherence takes place.

Our contribution to solve this problem is based in old ideas of Segal (\cite
{Segal}, \cite{Bogo}) and van Howe \cite{van Howe}, reformulated by Antoniou
et al. \cite{Antoniou}. We have developed these ideas in papers \cite{LauraA}
and \cite{LauraE} where we have shown how the Riemann-Lebesgue theorem can
be used to prove the destructive interference of the off-diagonal terms of
the state density matrix yielding decoherence. Using this technique we have
found decoherence and the classical statistical equilibrium limit in simple
quantum systems \cite{LauraF} where we have defined the final pointer basis
in an unambiguous way. \footnote{%
The relation of our method with the histories decoherence is studied in
paper \cite{LauraF}. They turn out to be equivalent, but in our method the
final pointer basis is more properly defined.}. Also the localization
phenomenon appears in some cases.

This paper is devoted to give two examples of the method introduced in
papers \cite{LauraA} and \cite{LauraE}, that we will briefly review in
section II, and used it to find a general solution for the quantum-classical
limit in paper \cite{LauraF}.

In section III, the method will be used to solve the problem known as ''the
Mott problem'' after the name of Sir Neville Mott, probably the first one
who studied the subject \cite{Mott}. Let us consider a radioactive nucleus,
placed at the origin of coordinates $O$, inside a bubble chamber. We will
see that classical radial trajectories appear in the chamber due to the
emitted out-going particles. We must explain this phenomenon. Theoretically
we have a timeless structure, since the wave function, $\psi ({\bf x)}$
satisfies the eigenvalue equation: 
\begin{equation}
H\psi =\omega \psi  \label{1.1}
\end{equation}
where: 
\begin{equation}
H=-\frac 1{2M}\frac 1{r^2}\frac \partial {\partial r}r^2\frac \partial {%
\partial r}+\frac{L^2}{2Mr^2}+W(r)  \label{1.2}
\end{equation}
being $W(r)$ the spherically symmetric potential barrier, which is the
external wall of a potential well, and such that $\lim_{r\rightarrow \infty
}W(r)=0.$ This will be our model for the nuclear forces \cite{BGM}. Let us
observe that there is no trace of the time in these equations.

The main problem is that, even if the symmetry is a spherical one, there is
no a priori reason to explain why the classical trajectories are radial.
Really we have the following list of facts to explain:

i.-Why the nucleus quantum regime becomes the classical regime for the
classical trajectories

ii.- Why these classical trajectories are radial.

iii.- Why the notion of time, necessary to explain the motion of the
classical radial particles, appears in a timeless formalism, i. e. in the
one of eq. (\ref{1.1}).

iv.-Why there are only outgoing motions \footnote{%
But let us complete the first example discussing the role of the global
atmosphere of the bubble chamber in our problem. This role is non essential
and really just incidental. In fact, if there were no bubble chamber, at
least for $r\rightarrow \infty $ the classical trajectories would also be
radial, since in this case, any small detector located far enough from the
origin would find radial motions. Furthermore it would be very difficult to
believe that these radial motions are produced only by the small detector.
Then the radial motion exists even if there is no bubble chamber and it is a
mistake to consider that it is the bubble chamber that, acting as an
environment, produces the radial structure. But we must remark that there is
a measurement in both cases and therefore the measurement processes is
essential.}.

At this stage we must observe that this set of problems is very similar to
the one of Quantum Cosmology, where, in fact, we must explain the outcome of
the classical regime, the appearance of time, the nature of the classical
trajectories in superspace, and the direction of the corresponding motion:
i. e. the arrow of time. Then several of the most important quantum universe
problems, that we will discuss in the second example, are already contained
in our humble Mott model,

So, in section IV, we will consider our second example: the quantum
cosmology problem, since the appearance of a classical universe in quantum
gravity models is the cosmological version of the first problem. Then,
decoherence must also appear in the universe \cite{Cast decoher}.

In this paper, using our method, we will solve the two examples and we will
find:

i.-Decoherence in all the dynamical variables and in a well defined final
pointer basis.

ii.-A final classical equilibrium limit, when $\hbar \rightarrow 0$, in such
a way that the Wigner function $F_{*}^W=\rho _{*}^{(cl)}$ of the asymptotic
diagonal matrix $\rho _{*}$ can be expanded as (e. g., in the cosmological
problem): 
\begin{equation}
\rho _{*}^{(cl)}([{\bf x}],[{\bf k}])=\int p_{\{l\}[{\bf a}]}\rho _{\{l\}[%
{\bf a}_0]}^{(cl)}([{\bf x}],[{\bf k}])d\{l\}d[{\bf a}]  \label{1.3}
\end{equation}
where $\rho _{\{l\}[{\bf a}_0]}^{(cl)}([{\bf x}],[{\bf k}])$ is a classical
density strongly peaked \footnote{%
Precisely: peaked as allowed by the uncertainty principle.} in a trajectory
defined by the initial coordinates ${\bf a}$ and the momenta $l$ and $%
p_{\{l\}[{\bf a}]}$ is the probability of each trajectory. As essentially
the limit of quantum mechanics is not classical mechanics but classical
statistical mechanics this is our final result: the density matrix is
translated in a classical density, via a Wigner function, and it is
decomposed as a sum of densities peaked around all possible classical
trajectories, each one of these densities weighted by their own probability 
\footnote{%
After classical statistical regime is reached correlations will eventually
produce a pure classical regime.}.

Thus our quantum density matrix behaves in its classical limit as a
statistical distribution among a set of classical trajectories. Similar
results, for the cosmological case, are obtained in papers \cite{Zouppas}
and \cite{Polarsky}.

We will reelaborate these conclusions in section V.

\section{Review of the method.}

In order to go from the quantum to the classical statistical regime two new
properties must appear:

i.-{\bf Decoherence:} The density matrices, that contain quantum
interference terms, must become diagonal, in such a way that these
interferences would be suppressed. Then the quantum way to find
probabilities of exhaustive events (i. e.: adding the corresponding
amplitude and computing the norm) become the classical way: just adding the
probabilities.

ii.-{\bf The limit }$\hbar \rightarrow 0${\bf :} The positions and the
momenta (or more generally canonically conjugated dynamical variables) can
be defined as allowed by the uncertainty principle, but big scales (i. e.
when{\bf \ }$\hbar \rightarrow 0)$ allow us to consider both the position
and the momentum as independent dynamical variables, like in classical
mechanics. Of course this independence is the essential property to find a
classical behavior.

These two closely related properties introduce the classical statistical
behavior in the quantum formalism. Let us begin with decoherence.

We are only interested in the scattering states with continuous spectrum, e.
g., for the first example, the Mott problem, the radial outgoing particles
are described by these states. To obtain the hamiltonian eigenbasis of the
Hilbert space ${\cal H}$ we can consider e. g. the hamiltonian (\ref{1.2})
and construct Lippmann-Schwinger basis $\{|\omega +\rangle \}$ \cite{Bohm},
(really $|\omega ,l,n+\rangle )$ where $\omega $ is partially discrete and
partially continuous, e. g. it will have some $\omega _0$ for the ground
state and a continuous $\omega $ for the scattering states, but for the
moment we will consider only the continuous index $\omega ,$ (since we are
only interested in these states \footnote{%
Bound states are considered in papers \cite{LauraA} and \cite{LauraE}.}$)$.
Of course we can as well use $\{|\omega ->\}.$ Then the hamiltonian can be
diagonalized as: 
\begin{equation}
H=\int_0^\infty \omega |\omega +><\omega +|d\omega  \label{2.4}
\end{equation}
From this expression we can deduce that the most general observable that we
can consider reads: 
\begin{equation}
O=\int_0^\infty O_\omega |\omega +><\omega +|d\omega +\int_0^\infty
\int_0^\infty O_{\omega \omega ^{\prime }}|\omega +><\omega ^{\prime
}+|d\omega d\omega ^{\prime }  \label{2.5}
\end{equation}
where the functions $O_\omega ,O_{\omega \omega ^{\prime }}$ are regular,
namely, the most general observable must have a singular component (the
first term of the r.h.s. of the last equation) and a regular part (the
second term). If the singular term would be missing the hamiltonian would
not belong to the space of the chosen observables \cite{LauraA}. Then $O\in 
{\cal O\subset H\oplus (H\otimes H)}$ since some extra conditions must be
added \cite{LauraA}. This space has the basis $\{|\omega ),|\omega ,\omega
^{\prime })\}$: 
\begin{equation}
|\omega )=|\omega +><\omega +|,\text{ }|\omega ,\omega ^{\prime })=|\omega
+><\omega ^{\prime }+|  \label{2.5'}
\end{equation}
The regular quantum state $\rho $ are measured by the observables just
defined computing the mean values of these observable in the quantum states $%
<O>_\rho =Tr(\rho O)$ \cite{Ballentine}. These mean values can be considered
as linear functionals $\rho $ on the vectors $O.$ Then the notion of state
can be generalized to any linear functional over ${\cal O}$ that we can call 
$(\rho |O)$ \cite{Bogo}$.$ In this way not only regular states but singular
states can be defined. Moreover, as $\rho $ must be normalized, selfadjoint
and positive definite, $\rho \in {\cal S\subset O}^{\prime },$ where ${\cal S%
}$ is a convex set contained in ${\cal O}^{\prime }$ \cite{LauraA}, \cite
{LauraE}. The basis of ${\cal O}^{\prime }$ is $\{(\omega |,(\omega ,\omega
^{\prime }|\}$. These states are defined as functionals by the equations: 
\begin{equation}
(\omega |\omega ^{\prime })=\delta (\omega -\omega ^{\prime }),\ (\omega
,\omega "|\omega ^{\prime },\omega ^{\prime \prime \prime })=\delta (\omega
-\omega ^{\prime })\delta (\omega "-\omega ^{\prime \prime \prime })
\label{2.5"}
\end{equation}
Therefore a generic quantum state reads: 
\begin{equation}
\rho =\int_0^\infty \rho _\omega (\omega |d\omega +\int_0^\infty
\int_0^\infty \rho _{\omega \omega ^{\prime }}(\omega ,\omega ^{\prime
}|d\omega d\omega ^{\prime }  \label{2.6}
\end{equation}
where $\rho _\omega \geq 0,\rho _{\omega \omega ^{\prime }}=\rho _{\omega
^{\prime }\omega }^{*}.$ The states such that $\rho _\omega \neq \rho
_{\omega \omega }$ will be called {\it generalized states }\cite{LauraA} and
those such that $\rho _\omega =\rho _{\omega \omega }$ are the usual regular
mixed or eventually pure states. To continue, even if the time is not
strictly defined (since we have only the eq. (\ref{1.1})) let us postulate
that there is a symmetry in the system with a symmetry group $e^{-iHt}$. At
this level of our reasoning this is a global fact imposed by the structure
of the universe where we suppose the model is immersed (we will come back to
this problem in subsection IIIC1, of course this will also be the case in
the cosmological example, because the problem of time definition is the
same). Then the time evolution of the quantum state $\rho $ reads: 
\begin{equation}
\rho (t)=\int_0^\infty \rho _\omega (\omega |d\omega +\int_0^\infty
\int_0^\infty \rho _{\omega \omega ^{\prime }}e^{i(\omega -\omega ^{\prime
})t}(\omega ,\omega ^{\prime }|d\omega d\omega ^{\prime }  \label{2.7}
\end{equation}
As in the statistical level we are considering we only measure mean values
of observables in quantum states, i. e.: 
\[
<O>_{\rho (t)}=(\rho (t)|O)= 
\]
\begin{equation}
\int_0^\infty \rho _\omega O_\omega d\omega +\int_0^\infty \int_0^\infty
\rho _{\omega \omega ^{\prime }}O_{\omega \omega ^{\prime }}e^{-i(\omega
-\omega ^{\prime })t}d\omega d\omega ^{\prime }  \label{2.8}
\end{equation}
using the Riemann-Lebesgue theorem we obtain the weak limit, for all $O\in 
{\cal O},$ $\rho \in {\cal S}$: 
\begin{equation}
\lim_{t\rightarrow \infty }<O>_{\rho (t)}=<O>_{\rho _{*}}  \label{2.9}
\end{equation}
where we have introduced the diagonal equilibrium state: 
\begin{equation}
\rho _{*}=\int_0^\infty \rho _\omega (\omega |d\omega  \label{2.10}
\end{equation}
Therefore, in a weak sense we have: 
\begin{equation}
W\lim_{t\rightarrow \infty }\rho (t)=\rho _{*}  \label{2.11}
\end{equation}
Thus, any quantum state goes to a equilibrium diagonal state weakly, and
that will be the result, if we observe and measure the system evolution with 
{\it any possible observable of space ${\cal O}$, e. g.: }for the first
example, with the global bubble chamber (or with any one of the local small
detectors of the footnote of the introduction){\it .} Then, from the
observational point of view we have decoherence of the energy levels, even
if, from the strong limit point of view the off-diagonal terms never vanish,
they just oscillate (eq. (\ref{2.7})). So, from now on, as we will consider
the matrix $\rho (t)$ for big $t$, e.g.: for the first example, far away
from the nucleus, the relevant state will be $\rho _{*}$, a diagonal state
in the energy.

Some observations are in order:

i.- The real existence of the two singular parts introduced above is assured
by the nature of the problem. The singular part of the observables is just a
necessary generalization of the singular part of the hamiltonian, which is
completely singular (eq. (\ref{2.4})). The singular part of the states is
the final state for decoherence (eq. (\ref{2.10})). $(\rho |O)$ is just the
natural generalization to the continuous of the trace of the product of two
finite dimensional matrices (eq. (\ref{2.8})).

ii.- To fully understand the phenomenon it is necessary to use generalized
states. Therefore, any explanation only based on pure wave function states
or mixed states is incomplete \footnote{%
Now an incidental question for the first example would be: which energies?
In the first example the potential well, surrounded by the barrier, $W(r)$
originates unstable levels with energy $\omega _n$ which decay with a
decaying time $\gamma _n^{-1}.$ Inside the well there are oscillating waves
that can be used to fulfill the boundary conditions at $r=0.$ When these
waves arrive to the barrier they are partially reflected and transmitted 
\cite{BGM}. The states that tunnel the barrier appear with an energy that
peaks strongly at $\omega _n$ \cite{LauraA}, \cite{CyLI}. Therefore the
energy of the outgoing particles are energy packets $\sim \delta (\omega
-\omega _n)$, and therefore could be essentially labelled with $n.$ But
taking into account that the energy spectrum is really a continuous one we
will always refer to it as $\omega .$}.

Having established the decoherence in the energy we must consider the
decoherence in the other dynamical variables. Before going to the model let
us study the general case (we will repeat this reasoning in sections IIB and
IVB, with the notation corresponding to each case). The diagonal singular
component of the eq. (\ref{2.7}) (which is equal to $\rho _{*})$ is time
independent, therefore it is impossible that a different decoherence process
takes place in this component to eliminate the off-diagonal terms, in the
other dynamical variables. Therefore, the only thing to do is to try to find
if there is a basis where these diagonal terms vanish at any time and
therefore there is a perfect and complete decoherence. For $t\rightarrow
\infty $ this basis in fact exists and it is known as the {\it final pointer
basis.}

Let $\{H,O_1,...O_N\}$ be the usual complete set of commuting observables
(CSCO) that we are using to make our calculations and $\{|\omega
,m_1,...,m_N+>\}$ (that we will simply call $\{|\omega ,m_1,...,m_N>\}$ from
now on) the corresponding eigenbasis. Then introducing the new indices in
eq. (\ref{2.10}) the equilibrium diagonal state reads: 
\begin{equation}
\rho _{*}=\int \sum_{m_1,...,m_N,m_1^{\prime },...,m_N^{\prime }}\rho
_{m_1,...,m_N,m_1^{\prime },...,m_N^{\prime }}^{(\omega )}(\omega
,m_1,...,m_N,m_1^{\prime },...,m_N^{\prime }|d\omega  \label{2.12}
\end{equation}
From what we have said under eq. (\ref{2.6}) it is: 
\begin{equation}
(\rho _{m_1,...,m_N,m_1^{\prime },...,m_N^{\prime }}^{(\omega )})^{*}=\rho
_{m_1^{\prime },...,m_N^{\prime },m_1,...,m_N}^{(\omega )}  \label{2.13}
\end{equation}
therefore this matrix can be diagonalized and there is a basis $\{(\omega
,l_1,...,l_N|\}$ where the matrix $\rho _{*}$ reads: 
\begin{equation}
\rho _{*}=\int \sum_{l_1,...,l_N}\rho _{l_1,...,l_N}^{(\omega )}(\omega
,l_1,...,l_N|d\omega  \label{2.14}
\end{equation}
Now we can define the observables: 
\begin{equation}
P_i=\int \sum_{l_1,...,l_N}P_{l_1,...,l_N}^{(i,\omega )}(\omega
,l_1,...,l_N|d\omega  \label{2.15}
\end{equation}
and the CSCO $\{H,P_1,...,P_N\},$ where the singular component $\rho _{*}$
is diagonal in the dynamical variables corresponding to the observables $%
P_1,...,P_N$ from the very beginning. This is the {\it final pointer basis}
where there is perfect decoherence. Final pointer basis is therefore defined
by the dynamics of the model and by the quantum state considered, in
complete agreement with the literature on the subject. The classical
statistical limit will be complete if we transform all these equations via a
Wigner integral as we will in each example using the corresponding notation.

\section{The Mott problem.}

\subsection{Decoherence.}

After these general considerations let us now go to our first problem. The
hamiltonian of eq. (\ref{1.2}) can be decomposed as: 
\begin{equation}
H=H_0+V  \label{2.1}
\end{equation}
\begin{equation}
H_0=-\frac 1{2M}\frac 1{r^2}\frac \partial {\partial r}r^2\frac \partial {%
\partial r}+\frac{L^2}{2Mr^2}  \label{2.2}
\end{equation}
\begin{equation}
V=W(r)  \label{2.3}
\end{equation}
allowing the definition of the Lippmann-Schwinger basis.

Its essential property is its spherical symmetry. We will see how this
symmetry leads directly to the result above, avoiding the diagonalization
procedure used in section II. In order to conserve this symmetry it is
necessary that the nucleus prepares only spherically symmetric states. Thus,
we can foresee that the CSCO corresponding to the pointer basis must contain
the generator of angular rotation so it must be $\{H,L^2,L_z\}.$ The pointer
basis must then be the usual $\{|\omega ,l,m>\}$ basis.

In fact, let us consider an initial state functional $\rho _0$ with
spherical symmetry. If $\overline{{\Bbb L}}$ is the generator the of
densities rotations, a rotation of the state by an angle $\overline{\varphi }
$ gives $\rho _0=\exp \{-i\,\overline{{\Bbb L}}\cdot \overline{\varphi }%
\}\,\rho _0$, or equivalently 
\[
(\exp \{-i\,\overline{{\Bbb L}}\cdot \overline{\varphi }\}\,\rho
_0\,|O)=(\rho _0|\exp \{i\,\overline{L}\cdot \overline{\varphi }\}\,O\,\exp
\{-i\,\overline{L}\cdot \overline{\varphi }\}), 
\]
for any observables $O\in {\cal O}$. In the last expression, $\overline{L}$
is the angular momentum operator. From the last equation we obtain 
\begin{equation}
(\overline{{\Bbb L}}\,\rho _0|O)=(\rho _0|[\overline{L},O])=0.  \label{R1}
\end{equation}

The observables $O$ have the form 
\begin{eqnarray}
O &=&\sum_{l\;m}\sum_{l^{\prime }\;m^{\prime }}\int d\omega
\,O_{lm,l^{\prime }m^{\prime }}(\omega )\,|\omega lm\rangle \langle \omega
l^{\prime }m^{\prime }|+  \nonumber \\
&&+\sum_{l\;m}\sum_{l^{\prime }\;m^{\prime }}\int \int d\omega \,d\omega
^{\prime }\,O_{lm,l^{\prime }m^{\prime }}(\omega ,\omega ^{\prime
})\,|\omega lm\rangle \langle \omega ^{\prime }l^{\prime }m^{\prime }|,
\label{R2}
\end{eqnarray}
and therefore equation (\ref{R1}) gives 
\begin{equation}
(\rho _0|[\overline{L},|\omega lm\rangle \langle \omega ^{\prime }l^{\prime
}m^{\prime }|])=0.  \label{X.1}
\end{equation}
These equations are equivalent to 
\begin{equation}
(\rho _0|[L_z,|\omega lm\rangle \langle \omega ^{\prime }l^{\prime
}m^{\prime }|])=0,\quad (\rho _0|[L_{\pm },|\omega lm\rangle \langle \omega
^{\prime }l^{\prime }m^{\prime }|])=0,\quad L_{\pm }=L_x\pm L_y.  \label{X.2}
\end{equation}
Taking into account that 
\begin{equation}
L_z|l,m\rangle =m|l,m\rangle \quad L_{\pm }|l,m\rangle =\sqrt{l(l+1)-m(m\pm
1)}\,|l,m\pm 1\rangle ,  \label{X.3}
\end{equation}
we obtain 
\[
0=(m-m^{\prime })(\rho _0||\omega ,l,m\rangle \langle \omega ^{\prime
},l^{\prime },m^{\prime }|) 
\]
\begin{eqnarray}
0 &=&\sqrt{l(l+1)-m(m\pm 1)}\;(\rho _0||\omega ,l,m\pm 1\rangle \langle
\omega ^{\prime },l^{\prime },m^{\prime }|)  \label{X.4} \\
&&-\sqrt{l^{\prime }(l^{\prime }+1)-m^{\prime }(m^{\prime }\mp 1)}\;(\rho
_0||\omega ,l,m\rangle \langle \omega ^{\prime },l^{\prime },m^{\prime }\mp
1|).  \nonumber
\end{eqnarray}
both for $\omega =\omega ^{\prime }$ or $\omega \neq \omega ^{\prime }.$
These equations give 
\begin{eqnarray}
(\rho _0||\omega ,l,m\rangle \langle \omega ^{\prime },l^{\prime },m^{\prime
}|) &=&\rho _l^0(\omega ,\omega ^{\prime })\;\delta _{l\;l^{\prime
}}\;\delta _{m\;m^{\prime }}  \label{X.5} \\
(\rho _0||\omega ,l,m\rangle \langle \omega ,l^{\prime },m^{\prime }|)
&=&\rho _l^0(\omega )\;\delta _{l\;l^{\prime }}\;\delta _{m\;m^{\prime }}. 
\nonumber
\end{eqnarray}
Thus any symmetric $\rho _0$ is diagonal, in $l$ and $m.$ We will repeat the
conclusion of the previous section and see, in this particular case, how the
time evolution produces the diagonalization: namely $\rho _l^0(\omega
,\omega ^{\prime })\,$ will vanishes when $t\rightarrow \infty .$ As in eq. (%
\ref{2.8}), the time evolution reads: 
\begin{eqnarray}
\langle O\rangle _{\rho _t} &=&(\rho _t|O)=(\rho _0|e^{iHt}Oe^{-iHt})=
\label{X.7} \\
&=&\int d\omega \sum_{l=0}^\infty \rho _l^0(\omega
)\sum_{m=-l}^{+l}O_{lm,lm}(\omega )+  \nonumber \\
&&+\int d\omega \int d\omega ^{\prime }\sum_{l=0}^\infty \rho _l^0(\omega
,\omega ^{\prime })\,\exp \{i(\omega -\omega ^{\prime
})t\}\sum_{m=-l}^{+l}O_{lm,lm}(\omega ,\omega ^{\prime }).  \nonumber
\end{eqnarray}
Riemann-Lebesgue theorem can be used in this expression to obtain the
''final'' state 
\begin{equation}
\langle O\rangle _{\rho _{*}}=(\rho _{*}|O)=\lim_{t\rightarrow \infty }(\rho
_t|O)=\int d\omega \sum_{l=0}^\infty \rho _l^0(\omega
)\sum_{m=-l}^{+l}O_{lm,lm}(\omega ).  \label{X.8}
\end{equation}
where $\rho _l^0(\omega ,\omega ^{\prime })\,$ has disappeared. If we
define, in analogy to eqs. (\ref{2.5'}) and (\ref{2.5"}) the functional $%
(\omega ,l,m|$ acting on an observable $O$ of the form given in equation (%
\ref{R2}) by $(\omega ,l,m|O)=O_{lm,lm}(\omega )$, we can give the following
expression for the asymptotic form of the state 
\begin{equation}
(\rho _{*}|=W\lim_{t\rightarrow \infty }(\rho _t|O)=\int d\omega
\sum_{l=0}^\infty \rho _l^0(\omega )\sum_{m=-l}^{+l}(\omega ,l,m|.
\label{X.9}
\end{equation}

Of course $(\rho _{*}|$ is also spherically symmetric since it is symmetric
in $l$ and $m$. We do not address in this paper the mechanism used by the
nucleus to prepare the states in such a spherically symmetric way. But we
are just studying the case where all the elements of the nucleus are
spherically symmetric, and also the quantum states involved, because we are
precisely trying to explain the breaking of this symmetry and the appearance
of the radial structure. Then, it is clear that the only possibility is to
begin with a spherically symmetric structure and with states satisfy the
equations above.

Therefore:

i.-The origin of the decoherence in the energy is the time evolution.

ii.- The origin of the decoherence in the angular variables is the
preparation of the quantum state, which is spherically symmetric in our
model. This symmetry is preserved by the time evolution.

Let us observe that, as our model is spherically symmetric, any CSCO $\{H,%
{\bf L}^2,L_z\},$ for any arbitrary $z$ axis, will correspond to the final
pointer basis. But if the symmetry would be cylindrical along the axis $z,$
being the center of an angular coordinate $\varphi $ with generator $L_z $,
the only CSCO related with a final pointer basis, for the cylindrical
quantum states, would be $\{H,p_z,L_z\}.$ So we see how the symmetry of the
equation and the states defines the final pointer bases and their number.

\subsection{The limit $\hbar \rightarrow 0$ and the classical $\rho
_{*}^{(cl)}(q,p).$}

Let us now compute the classical analogue of $\rho _{*},$ as promised at the
end of section II. We will prove that the distribution function $\rho
_{*}^{(W)}(q,p)$, that corresponds to the density matrix $\rho _{*}$ via the
Wigner integral \cite{Wigner} is simply a function of the classical constant
of the motion, in our case $H(q,p),$ ${\bf L}^2(q,p),$ $L_z(q,p),$
precisely: 
\begin{equation}
\rho _{*}^{(W)}(q,p)=\rho _{*}(H(q,p),{\bf L}^2(q,p),L_z(q,p))  \label{4.1}
\end{equation}
To simplify the demonstration let us consider only the constant $H(q,p).$
From eq. (\ref{2.10}) we have:

\begin{equation}
\rho _{*}=\int \rho _{*}(\omega )(\omega |d\omega  \label{4.2}
\end{equation}
So we must compute: 
\begin{equation}
\rho _\omega ^{(W)}(q,p)=\pi ^{-1}\int (\omega ||q+\lambda \rangle \langle
q-\lambda |)e^{2ip\lambda }d\lambda  \label{4.3}
\end{equation}
We know, from \cite{LauraA}, section II C, that the characteristic property
of $(\omega |$ is: 
\begin{equation}
(\omega |H^n)=\omega ^n  \label{4.4}
\end{equation}
Using the relation between quantum and classical inner products of operators
(\cite{Wigner} eq. (2.13)) we deduce that the characteristic property of $%
\rho _\omega ^{(W)}(q,p)$ is: 
\begin{equation}
\int \rho _\omega ^{(W)}(q,p)[H(q,p)]^ndqdp=\omega ^n+O(\hbar )  \label{4.5}
\end{equation}
for any natural number $n.$ From now on we will work in the limit $\hbar
\rightarrow 0$ and therefore all the $O(\hbar )$ will disappear. Thus $\rho
_\omega ^{(cl)}(q,p)$ must be: 
\begin{equation}
\rho _\omega ^{(W)}(q,p)=\delta (H(q,p)-\omega )>0  \label{4.6}
\end{equation}
which turns out to be a distribution, namely a functional as $\rho _{*}$.
Therefore, going back to eq. (\ref{4.2}) and since the Wigner relation is
linear, we have: 
\begin{equation}
\rho _{*}^{(W)}(q,p)=\int \rho _{*}(\omega )\rho _\omega ^{(W)}(q,p)d\omega
=\int \rho _{*}(\omega )\delta (H(q,p)-\omega )d\omega =\rho _{*}(H(q,p))>0
\label{4.7}
\end{equation}
q.e.d.

Generalizing this reasoning we can prove eq. (\ref{4.1}). Moreover the
generalized eq. (\ref{4.7}) reads: 
\begin{equation}
\rho _{*}^{(W)}(q,p)=\int \sum_{l,m}\rho _{*}(\omega ,l,m)\rho _{\omega
,l,m}^{(W)}(q,p)d\omega  \label{4.7'}
\end{equation}
where $\rho _{*}(\omega ,l,m)=\rho _l^0(\omega )$ of eq. (\ref{X.5}) 
\footnote{%
Even if, by symmetry reasons $m$ is absent as an index of $\rho _l^0(\omega
) $ we have introduced this index in $\rho _{*}(\omega ,l,m)$ for two
reasons:
\par
i.-It may be present in a more general case (as the cosmological one).
\par
ii.- It is present in $\rho _{\omega ,l,m}^{(cl)}.$} and $\rho _{\omega
,l,m}^{(W)}(q,p)$ reads: 
\begin{equation}
\rho _{\omega ,l,m}^{(W)}(q,p)=\delta (H(q,p)-\omega )\delta ({\bf L}%
^2(q,p)-l(l+1))\delta (L_z(q,p)-m)  \label{4.8}
\end{equation}
and can be interpreted as the state where $\omega ,$ $l,$ $m$ are well
defined and the corresponding classical canonically conjugated variables
completely undefined since $\rho _{\omega ,l,m}^{(W)}$ is not a function of
these variables. Then there is a complete coincidence with the result of the
previous section. From the last two equations we obtain (\ref{4.1}) as we
promised. Now all the classical canonically conjugated variables of the
momenta $H,$ ${\bf L}^2,$ $L_z$ do exist since they can be found solving the
corresponding Poisson brackets differential equations \footnote{%
On the contrary, $H$ and $L^2$ have not quantum canonically conjugated
dynamical variables since their spectra are bounded for below.}. But as the
momenta $H,{\bf L}^2,L_z$, that we will call generically $l,$ are also
constants of the motion, we have $l^{\bullet }=-\partial H/\partial a=0$,
where $a$ is the coordinate canonically conjugated to $l,$ so $H$ is just a
function of $l$ and: 
\begin{equation}
a^{\bullet }=\frac{\partial H(l)}{\partial l}=\varpi (l)=const.  \label{4.9}
\end{equation}
so: 
\begin{equation}
a=\varpi (l)t+a_0  \label{4.10}
\end{equation}
These are the classical motions corresponding to the motion of the wave
packet of the previous subsection. As in this section $l$ is completely
defined and $a_0$ completely undefined due to spherical symmetry, in such a
way that the motions represented in the last equation homogeneously fill the
surface $l=const$. (really $H=const,$ ${\bf L}^2=const.,$ $L_z=const.$ for
our case), namely a usual classical torus of phase space.

Then, eq. (\ref{4.7'}) can be considered as the expansion of $\rho
_{*}^{(cl)}(q,p)$ in the classical motion just described, contained in $\rho
_{n,l,m}^{(cl)}(q,p),$ each one with a probability $\rho _{*}(n,l,m).$

Summing up:

i.- We have shown that the density matrix $\rho (t)$ evolves to a diagonal
density matrix $\rho _{*.}$

ii.- This density matrix has $\rho _{*}^{(cl)}(q,p)$ as its corresponding
classical density.

iii.- This classical density can be decomposed in classical motions where $%
H, $ ${\bf L}^2,$ and $L_z$ remains constant, answering question i of the
introduction.

iv.- Finally, for $r\rightarrow \infty $ the classical analogue of eq. (\ref
{2.1}): 
\begin{equation}
H=\frac 1{2M}p_r^2+\frac{{\bf L}^2}{2Mr^2}+W(r)  \label{4.11}
\end{equation}
show that in the classical motions we can consider ${\bf L}^2=0$ when $%
r\rightarrow \infty $ \footnote{%
A close study of the initial conditions, that necessarily are located near
the center of the nucleus will improve the demonstration of this result. For
the sake of conciseness we do not include these reasonings.}$.$ Thus the
classical motions far from the nucleus can be considered as radial,
answering question ii of the introduction.

Finally as: 
\begin{equation}
\int \delta (\theta -\theta ^{(0)})\delta (\varphi -\varphi ^{(0)})d\theta
^{(0)}d\varphi ^{(0)}=1  \label{4.12}
\end{equation}
where $\theta $ and $\varphi $ are the usual polar coordinates. Since far
from the nucleus $l=m=0,$ we can write eq. (\ref{4.7'}) as: 
\begin{equation}
\rho _{*}^{(W)}(q,p)=\int \rho _{*}(\omega ,0,0)\delta (H(q,p)-\omega
)\delta (\theta -\theta ^{(0)})\delta (\varphi -\varphi ^{(0)})d\theta
^{(0)}d\varphi ^{(0)}d\omega  \label{4.13}
\end{equation}
where $\delta (H(q,p)-\omega )\delta (\theta -\theta ^{(0)})\delta (\varphi
-\varphi ^{(0)})$ can be considered as the classical density of the
classical particles with radial motions defined by the energy $\omega $ and
the angles $\theta ^{(0)}$ and $\varphi ^{(0)}.$ The final classical
equilibrium density is thus decomposed in the densities of radial classical
trajectories, as in eq. (\ref{1.3}).

So finally we have an isotropic ensemble of particles in radial motion. This
is the classical statistical limit of our wave function. We have gone from
quantum mechanics to classical statistical mechanics \cite{Ballentine}. To
single out any one of the trajectories to obtain a single classical motion
is clearly impossible, at the statistical mechanics level, since it would
brake the spherical symmetry of the initial wave function. But the
statistical state is composed of single radial motions, at the classical
level, as any population is composed of individuals.

\subsection{Localization and correlations.}

These phenomena will appears for adequate potential and initial conditions
(see \cite{LauraF}). If $W(r)=0$ we know that the wave packet will spread,
so the real $W(r)$ must be such that the eventual spreading be as slow as
necessary in order to see the radial trajectories in the bubble chamber.

\subsection{Discussions and comments.}

\subsubsection{The time.}

We have postulate the global existence of time in section IIA. This
postulate was motivated in the fact that this is a global feature of the
universe. But, in order to mimic the cosmological case of the second
example, let us imagine for a moment that our model would be considered as a
model of the whole universe \footnote{%
This will be the case in the first problem. In the second one we will deal
with the real universe and reproduce the same kind of arguments.}. Then it
would be impossible to postulate the existence of time based in an exterior
global structure. A way to solve this problem would be to postulate
decoherence through eq. (\ref{2.10}), namely the existence of some kind of
evolution with such a final diagonal state which corresponds to the fact
that the universe really ends in a classical states. Then we can define the
time over the classical trajectory labelled by $\omega $ and $l$ via eq. (%
\ref{4.11}) as: 
\begin{equation}
M\frac{dr}{dt}=\sqrt{2M\left[ \omega -\frac{l(l+1)}{2Mr^2}-W(r)\right] }
\label{5.1}
\end{equation}
i. e.: 
\begin{equation}
t=\frac 1{\sqrt{2}}\int_0^r\frac{dr}{\sqrt{\omega -\frac{l(l+1)}{2Mr^2}-W(r)}%
}  \label{5.2}
\end{equation}
Thus we can see the interrelation of the time, the decoherence, and the
elimination of the uncertainty relations. The usual procedure is to
postulate the existence of time and follow the chain time$\rightarrow $%
decoherence$\rightarrow $elimination of the uncertainty relations. In this
case we are using the traditional way of thinking of classical and non
relativistic quantum mechanics: {\it time is a primitive concept.} But, as
we have explained, another chain is feasible: decoherence$\rightarrow $%
elimination of the uncertainty relations$\rightarrow $time, and that would
be the chain that we must use if we consider the whole universe. In this
last case we could postulate (based on a obvious observational fact) that
the quantum physics of the universe is such that it has a tendency towards a
classical regime (decoherence + elimination of the uncertainty relations).
This tendency would be the primitive concept in this case and {\it time
would be a derived concept} in perfect accord with Mach philosophy \cite
{Barbour} (see also \cite{Castagnino}). So question iii of the introduction
is answered. We will come back to these arguments in the cosmological case
in section IVD3 where we will go further on. We will postulate the existence
of a parameter $\eta ,$ that at the quantum level will take the role of
time, namely $|\eta \rangle =e^{-iHt}|0\rangle ,$ and show that there is
decoherence in this parameter that, therefore, becomes a classical one.

\subsubsection{Why the outgoing solutions?}

The answer to this question can be searched in one of the essential
properties of the actual state of the real universe: {\it it is
time-asymmetric}, in spite of the fact that its evolution laws are
time-symmetric, e. g.:

i.- Even if we have incoming and outgoing scattering states, only the second
ones behave spontaneously, while the first ones must be produced by a source
of energy \cite{Goslar} \footnote{%
Incoming states will transform the stable nucleus in a unstable one.
Outgoing states will radiate the energy produced when the unstable nucleus
evolves to a equilibrium state in an spontaneous way.}.

ii.- Even if we have advanced and retarded solutions of the Maxwell
equations the time asymmetry of the state of the universe force us to use
the later ones and to neglect the former ones.

iii.-At the quantum level, the space of admissible solutions is not the
whole Hilbert Space but a subspace of this space, where causality can be
introduced \footnote{%
As causality is related with the analytic properties of wave functions ,
when we promote the energy $\omega $ to a complex variable $z,$ we can
choose the admissible solutions as those wave functions that are analytic
and bounded in the lower complex half-plane of variable $z$, even if we
could choose those that have the same properties in the upper plane(or more
precisely such that they belong to the Hardy class function space from below
(or above) \cite{Bohm}). In this way causality is introduced and its
physical consequences: dispersion relation, the fluctuation-dissipation
theorem, the growing of entropy, etc. \cite{CyLI}, \cite{CyLIII}.}.

Namely, as the laws of physics are time symmetric they always give us two
t-symmetric possibilities. The universe is t-asymmetric and corresponds to
one of this possibilities.

From what we have said it must be clear that, in order to be completely
satisfactory, the choice must be done in the whole universe, namely it must
be global \cite{Goslar}$.$ Let us only consider the case ''i'', we must only
choose outgoing states answering question iv.

For our first problem we may use the WKB solution of the equation: 
\begin{equation}
\frac{d^2\psi (x)}{dx^2}+k^2(x)\psi (x)=0  \label{6.9}
\end{equation}
which is: 
\begin{equation}
\psi (x)=[k(x)]^{-\frac 12}\exp [\pm i\int k(x)dx]  \label{6.10}
\end{equation}
Then if we write the solutions of the eq. (\ref{1.2}) as:

\begin{equation}
\psi _{\omega lm}({\bf x})=\psi _{\omega lm}(r,\theta ,\varphi
)=Y_l^m(\theta ,\varphi )\frac{u_{\omega l}(r)}r  \label{6.1}
\end{equation}
the function $u_{\omega l}(r)$ satisfies the equation: 
\begin{equation}
\frac{d^2u_{\omega l}(r)}{dr^2}+2M\left[ \omega -\frac{l(l+1)}{2Mr^2}%
-W(r)\right] u_{\omega l}(r)=0  \label{6.11}
\end{equation}
If we can call: 
\begin{equation}
k_{\omega l}^2(r)=2M\left[ \omega -\frac{l(l+1)}{2Mr^2}-W(r)\right]
\label{6.12}
\end{equation}
the WKB solution is : 
\begin{equation}
u_{\omega l}(r)=[k_{\omega l}(r)]^{-\frac 12}\exp [\pm i\int_0^rk_{\omega
l}(r)dr]  \label{6.13}
\end{equation}

These are the two possibilities produced by our time symmetric theory. Then,
in order that solution (\ref{6.13}) would be out-going we must choose the
sign +. In fact, far from the nucleus, when $t\rightarrow \infty
,r\rightarrow \infty $ we have $k_{\omega l}(r)\rightarrow k_\omega =+\sqrt{%
2M\omega }$ and considering the time evolution factor we have: 
\begin{equation}
e^{-iHt}u_{\omega l}(r)=[k_\omega ]^{-\frac 12}\exp [-i(\omega t-k_\omega
r+const.)]  \label{6.14}
\end{equation}
and the wave evolution becomes the outgoing unilateral shift: 
\begin{equation}
r=\frac \omega {k_\omega }t+const.  \label{6.15}
\end{equation}
So the question iv of the introduction is also answered \footnote{%
Perhaps it is interesting to observe that, following the ideas of paper \cite
{Barbour}, this typical outgoing nature of the shift (that is extremely
important in the cosmological models \cite{Goslar}) it is only possible
because, from the very beginning the configuration space has a
characteristic structure: it is spherically symmetric and this fact {\it %
defines the asymmetry} $r>0$, namely the asymmetry that says that the origin 
$O$ is substantially different than the sphere at the infinity. Therefore: 
{\it it is an asymmetry in configuration space that really introduces
time-asymmetry }\cite{Barbour}. Without this asymmetry all the treatment
would be impossible, since the ''outgoing'' notion itself would be
meaningless, and the choice of the lower half-plane unmotivated, (at least
in the case where we consider our system as the whole universe and therefore
we have no other physical phenomena to play with). Therefore the fact that
an isolated nucleus radiates and never receives spontaneously radiation from
the exterior, in a conspirative way, is a consequence of the fundamental
time-asymmetry of the universe. But if, as a theoretical example, we
consider this isolated system as the whole universe it is a consequence of
the asymmetry of the configuration space of the model, in complete agreement
with ref. \cite{Barbour}. In the second example the time asymmetry has, more
or less, the same origin: the global asymmetry of the universe as explained
in reference \cite{Goslar}.}.

\section{The cosmological problem.}

\subsection{The model.}

Let us consider the flat Roberson-Walker universe (\cite{Juanpa}, \cite{Cast}%
) with a metric: 
\begin{equation}
ds^2=a^2(\eta )(d\eta ^2-dx^2-dy^2-dz^2)  \label{12.1}
\end{equation}
where $\eta $ is the conformal time and $a$ the scale of the universe. Let
us consider a free neutral scalar field $\Phi $ and let us couple this field
with the metric, with a conformal coupling ($\xi =\frac 16)$. The total
action reads $S=S_g+S_f$ $+S_i$ and the gravitational action is: 
\begin{equation}
S_g=M^2\int d\eta [-\frac 12\stackrel{\bullet }{a}^2-V(a)]  \label{12.2}
\end{equation}
where $M$ is the Planck mass, $\stackrel{\bullet }{a}=da/d\eta ,$ and the
potential $V$ contains the cosmological constant term and eventually the
contribution of some form of classical mater. We suppose that $V$ has a
bounded support $0\leq a\leq a_1.$ We expand the field $\Phi $ as: 
\begin{equation}
\Phi (\eta ,{\bf x})=\int f_{{\bf k}}e^{-i{\bf k\cdot x}}d{\bf k}
\label{12.3}
\end{equation}
where the components of ${\bf k}$ are three continuous variables.

The Wheeler De-Witt equation for this model reads (compare with eq. (\ref
{1.1}) for the first example)): 
\begin{equation}
H\Psi (a,\Phi )=(h_g+h_f+h_i)\Psi (a,\Phi )=0  \label{12.4}
\end{equation}
where: 
\[
h_g=\frac 1{2M^2}\partial _a^2+M^2V(a) 
\]
\[
h_f=-\frac 12\int (\partial _{{\bf k}}^2-k^2f_{{\bf k}}^2)d{\bf k} 
\]

\begin{equation}
h_i=\frac 12m^2a^2\int f_{{\bf k}}^2d{\bf k}  \label{12.5}
\end{equation}
being $m$ the mass of the scalar field, ${\bf k}/a$ is the linear momentum
of the field, and $\partial _{{\bf k}}$ =$\partial /\partial f_{{\bf k}}.$

We can now go to the semiclassical regime using the WKB method (\cite{Hartle}%
), writing $\Psi (a,\Phi )$ as: 
\begin{equation}
\Psi (a,\Phi )=\exp [iM^2S(a)]\chi (a,\Phi )  \label{12.6}
\end{equation}
and expanding $S$ and $\chi $ as: 
\begin{equation}
S=S_0+M^{-1}S_1+...,\qquad \chi =\chi _0+M^{-1}\chi _1+...  \label{12.7}
\end{equation}
To satisfy eq. (\ref{12.4}) at the order $M^2,$ $S(a),$ the principal Jacobi
function, must satisfy the Hamilton-Jacobi equation: 
\begin{equation}
\left( \frac{dS}{da}\right) ^2=2V(a)  \label{12.8}
\end{equation}
We can now define the (semi)classical time as in section III C (but now
using an approximate solution only). It is a parameter $\eta =\eta (a)$ such
that: 
\begin{equation}
\frac d{d\eta }=\frac{dS}{da}\frac d{da}=\pm \sqrt{2V(a)}\frac d{da}
\label{12.9}
\end{equation}
So: 
\begin{equation}
\eta =\frac 1{\sqrt{2}}\int_{a_0}^a\frac 1{\sqrt{V(a)}}da  \label{12.9'}
\end{equation}
that must be compared with eq. (\ref{5.2}), but in this equation the
trajectories are completely classic, while in eq. (\ref{12.9'}) only $a$ is
classic while $\Phi $ remains as a quantum variable. The solution of eq. (%
\ref{12.9'}) is $a=\pm F(\eta ,C),$ where $C$ is an arbitrary integration
constant. Different values of this constant and of the $\pm $ sign give
different classical solutions for the geometry.

Then, in the next order of the WKB expansion, the Schroedinger equation
reads: 
\begin{equation}
i\frac{d\chi }{d\eta }=h(\eta )\chi  \label{12.10}
\end{equation}
where: 
\begin{equation}
h(\eta )=h_f+h_i(a)  \label{12.11}
\end{equation}
precisely: 
\begin{equation}
h(\eta )=-\frac 12\int \left[ -\frac{\partial ^2}{\partial f_{{\bf k}}^2}%
+\Omega _{{\bf k}}^2(a)f_k^2\right] d{\bf k}  \label{12.12}
\end{equation}
where: 
\begin{equation}
\Omega _\varpi ^2=m^2a^2+k^2=m^2a^2+\varpi  \label{12.13}
\end{equation}
and $\varpi =k^2$ and $k=|{\bf k|.}$ So the time dependence of the
hamiltonian comes from the function $a=a(\eta ).$

Let us now consider a scale of the universe such that $a_{out}\gg a_1$. We
will consider the evolution in this region where the geometry is almost
constant. Therefore we have an adiabatic final vacuum $|0\rangle $ and
adiabatic creation and annihilation operators $a_{{\bf k}}^{\dagger }$ and $%
a_{{\bf k}}.$ Then $h=h(a_{out})$ reads: 
\begin{equation}
h=\int \Omega _\varpi a_{{\bf k}}^{\dagger }a_{{\bf k}}d{\bf k}
\label{12.14}
\end{equation}

We can now consider the Fock space and a basis of vectors: 
\begin{equation}
|{\bf k}_1,{\bf k}_2,...,{\bf k}_n,...\rangle \cong |\{k{\bf \}\rangle =}a_{%
{\bf k}_1}^{\dagger }a_{{\bf k}_2}^{\dagger }...a_{{\bf k}_n}^{\dagger
}...|0\rangle  \label{12.15}
\end{equation}
where we have called $\{k{\bf \}}$ to the set ${\bf k}_1,{\bf k}_2,...,{\bf k%
}_n,...$ The vectors of this basis are eigenvectors of $h:$%
\begin{equation}
h|\{k{\bf \}\rangle =}\omega |\{k{\bf \}\rangle }  \label{12.16}
\end{equation}
where: 
\begin{equation}
\omega =\sum_{{\bf k\in \{k\}}}\Omega _\varpi =\sum_{{\bf k\in \{k\}}%
}(m^2a_{out}^2+\varpi )^{\frac 12}  \label{12.17}
\end{equation}
We can now use this energy to label the eigenvector as: 
\begin{equation}
|\{k{\bf \}\rangle =}|\omega ,[{\bf k]\rangle }  \label{12.18}
\end{equation}
where $[{\bf k]}$ is the remaining set of labels necessary to define the
vector unambiguously. $\{|\omega ,[{\bf k]\rangle \}}$ is obviously an
orthonormal basis so eq. (\ref{12.14}) reads: 
\begin{equation}
h=\int \omega |\omega ,[{\bf k]\rangle }\langle \omega ,[{\bf k]|}d\omega d[%
{\bf k]}  \label{12.19}
\end{equation}
This is the hamiltonian that corresponds to (\ref{2.4}) in the cosmological
case.

\subsection{Decoherence in the other dynamical variables.}

We can obtain the decoherence in the energy as in section II. Then, if we
reintroduce the other dynamical variables in eq. (\ref{2.10}) we obtain: 
\begin{equation}
(\rho _{*}|=\int \rho _{\omega [{\bf k][k}^{\prime }]}(\omega ,[{\bf k],[k}%
^{\prime }]|d\omega d[{\bf k]d[k}^{\prime }]  \label{14.1}
\end{equation}
where \{$(\omega ,[{\bf k],[k}^{\prime }]|,(\omega ,\omega ^{\prime },[{\bf %
k],[k}^{\prime }]\}$ is the cobasis \{$(\omega |,(\omega ,\omega ^{\prime
}|\}$ but now showing the hidden $[{\bf k].}$ This equation corresponds to (%
\ref{2.12}) in the cosmological case.

Let us observe that if we would use polar coordinates for ${\bf k}$ eq.(\ref
{12.3}) reads:

\begin{equation}
\Phi (x,n)=\int \sum_{lm}\phi _{klm}dk  \label{14.2}
\end{equation}
where:

\begin{equation}
\phi _{klm}=f_{k,l}(\eta ,r)Y_m^l(\theta ,\varphi )  \label{14.3}
\end{equation}
where $k$ is a continuous variable, $l=0,1,...,;$ $m=-l,...,l;$ and $Y$ are
spherical harmonic functions. So the indices $k,l,m$ contained in the symbol 
${\bf k}$ are partially discrete and partially continuous.

As $\rho _{*}^{\dagger }=\rho _{*}$ then $\rho _{\omega [{\bf k}^{\prime }%
{\bf ][k}]}^{*}=$ $\rho _{\omega [{\bf k][k}^{\prime }]}$ and therefore a
set of vectors $\{|\omega ,[{\bf l}]\rangle \}$ exists such that: 
\begin{equation}
\int \rho _{\omega [{\bf k][k}^{\prime }]}|\omega ,[{\bf l}]\rangle _{[{\bf k%
}^{\prime }]}d[{\bf k}^{\prime }]=\rho _{\omega [{\bf l]}}|\omega ,[{\bf l}%
]\rangle _{[{\bf k]}}  \label{14.4}
\end{equation}
namely \{$|\omega ,[{\bf l}]\rangle \}$ is the eigenbasis of the operator $%
\rho _{\omega [{\bf k][k}^{\prime }]}.$ Then $\rho _{\omega [{\bf l]}}$ can
be considered as an ordinary diagonal matrix in the discrete indices like
the $l$ and the $m$, and a generalized diagonal matrix in the continuous
indices like $k$ \footnote{{E. g.: We can deal with this generalized matrix
rigging the space ${\cal S}$ and using the Gel'fand-Maurin theorem \cite
{Gorini}, this procedure allows us to define a generalized state eigenbasis
for system with continuous spectrum. It has been used to diagonalize
hamiltonians with continuous spectra in \cite{CyLI}, \cite{Bohm}, \cite{CGG}%
, etc.}}{.} Under the diagonalization process eq. (\ref{14.1}) is written
as: 
\begin{equation}
(\rho _{*}|=\int U_{[{\bf k}]}^{\dagger [{\bf l}]}\rho _{\omega [{\bf l][l}%
^{\prime }]}U_{[{\bf k}^{\prime }]}^{[{\bf l}^{\prime }]}U_{[{\bf k}^{\prime
}]}^{\dagger [{\bf l}^{\prime \prime }]}(\omega ,[{\bf l}^{\prime \prime }%
{\bf ],[l}^{\prime \prime \prime }]|U_{[{\bf k}]}^{[{\bf l}^{\prime \prime
\prime }]}d\omega d[{\bf k]d[k}^{\prime }]d[{\bf l]d[l}^{\prime }]d[{\bf l}%
^{\prime \prime }{\bf ]d[l}^{\prime \prime \prime }]  \label{14.4'}
\end{equation}
where $U_{[{\bf k}]}^{\dagger [{\bf l}]}$ is the unitary matrix used to
perform the diagonalization and: 
\begin{equation}
\rho _{\omega [{\bf l][l}^{\prime }]}=\rho _{\omega [{\bf l]}}\delta _{[{\bf %
l][l}^{\prime }]}  \label{14.4''}
\end{equation}
where: 
\begin{equation}
\rho _{\omega [{\bf l][l}]}=\rho _{\omega [{\bf l]}}=\int U_{[{\bf l}]}^{[%
{\bf k}]}\rho _{\omega [{\bf k][k}^{\prime }]}U_{[{\bf l}]}^{\dagger [{\bf k}%
^{\prime }]}d[{\bf k}]d[{\bf k}^{\prime }]  \label{14.4'''}
\end{equation}
so we can define: 
\begin{equation}
(\omega ,[{\bf l]}|=(\omega ,[{\bf l],[l}]|=\int U_{[{\bf l}]}^{[{\bf k}%
]}(\omega ,[{\bf k],[k}^{\prime }]|U_{[{\bf l}]}^{\dagger [{\bf k}^{\prime
}]\dagger }d[{\bf k}]d[{\bf k}^{\prime }]  \label{14.4''''}
\end{equation}
We can repeat the procedure with vectors $(\omega ,\omega ^{\prime },[{\bf %
k],[k}^{\prime }]|$ and obtain vector $(\omega ,\omega ^{\prime },[{\bf l]|.}
$ In this way we obtain a diagonalized cobasis \{$(\omega ,[{\bf l]}%
|,(\omega ,\omega ^{\prime },[{\bf l]}\}.$ So we can now write the
equilibrium state as: 
\begin{equation}
\rho _{*}=\int \rho _{\omega [{\bf l]}}(\omega ,[{\bf l}]|d\omega d[{\bf l}]
\label{14.5}
\end{equation}
which corresponds to (\ref{2.14}) in the cosmological case. Since vectors $%
(\omega ,[{\bf l}]|$ can be considered as diagonals in all the variables we
have obtained decoherence in all the dynamical variables. This fact will
become clearer once we study the observables related with this vector and
introduce the notion of {\it final pointer basis.}

So, let us now consider the observable basis \{$|\omega ,[{\bf l]}),|\omega
,\omega ^{\prime },[{\bf l])}\}$ dual to the state cobasis $\{(\omega ,[{\bf %
l]}|,(\omega ,\omega ^{\prime },[{\bf l]|}\}.$ From eq. (\ref{2.5'}) and as
the $\omega $ does not play any role in the diagonalization procedure we
obtain: 
\begin{equation}
|\omega ,[{\bf l]})=|\omega ,[{\bf l}]\rangle \langle \omega ,[{\bf l}%
]|,\qquad |\omega ,\omega ^{\prime },[{\bf l])=}|\omega ,[{\bf l}]\rangle
\langle \omega ^{\prime },[{\bf l}]|  \label{14.5'}
\end{equation}
So in the basis \{$|\omega ,[{\bf l]}),|\omega ,\omega ^{\prime },[{\bf l])}%
\}$ the hamiltonian reads: 
\begin{equation}
h=\int \omega |\omega ,[{\bf l}])d\omega d[{\bf l}]=\int \omega |\omega ,[%
{\bf l}]\rangle \langle \omega ,[{\bf l}]|d\omega d[{\bf l}]  \label{14.6}
\end{equation}
Now, we can also define the operators: 
\begin{equation}
{\bf L}=\int {\bf l}|\omega ,[{\bf l}])d\omega d[{\bf l}]=\int {\bf l}%
|\omega ,[{\bf l}]\rangle \langle \omega ,[{\bf l}]|d\omega d[{\bf l}]
\label{14.7}
\end{equation}
that can also be written, as in eq. (\ref{2.15}): 
\begin{equation}
L_i=\int l_i|\omega ,[{\bf l}])d\omega d[{\bf l}]=\int l_i|\omega ,[{\bf l}%
]\rangle \langle \omega ,[{\bf l}]|d\omega d[{\bf l}]  \label{14.8}
\end{equation}
where $i$ is an index such that it covers all the dimension of the ${\bf l}$ 
\footnote{%
In principle the matter field $\Phi $ may have any number of particles $N$.
But since we are working in the final stage of the universe evolution with $%
a\sim a_{out}$, this number is a constant. Then the number of observables in
the CSCO is $4N$ and the ket in configuration variables read 
\mbox{$\vert$}%
$\eta ,[{\bf x}]\rangle =|\{x\}\rangle ,$ where $[{\bf x]}$ is the space
position of the $4N$ particles.}. Now we can consider the set $(h,L_i), $
which is a CSCO, since all the members of the set commute, because they
share a common basis and find the corresponding eigenbasis of the set,
precise $|\omega ,[{\bf l}]\rangle $ since \footnote{%
In some occasions we will call $h=L_0$ and $\omega =l_0.$}: 
\begin{equation}
h|\omega ,[{\bf l}]\rangle =\omega |\omega ,[{\bf l}]\rangle  \label{14.9}
\end{equation}
\begin{equation}
L_i|\omega ,[{\bf l}]\rangle =l_i|\omega ,[{\bf l}]\rangle  \label{14.10}
\end{equation}
Of course the $L_i$ are constant of the motion because they commute with $h.$
From all these equations we can say that:

i.- $(h,L_i)$ is the final pointer CSCO.

ii.- \{$|\omega ,[{\bf l]}),|\omega ,\omega ^{\prime },[{\bf l])}\}$ is the
final pointer observable basis.

iii.-$\{(\omega ,[{\bf l]}|,(\omega ,\omega ^{\prime },[{\bf l]|}\}$ is the
final pointer states cobasis.

In fact, from eq. (\ref{14.5}) we see that the final equilibrium state has
only diagonal terms in this state (those corresponding to vectors $(\omega ,[%
{\bf l]}|)$ , it has not off-diagonal terms (those corresponding to vectors $%
(\omega ,\omega ^{\prime },[{\bf l]|,}(\omega ,[{\bf k],[k}^{\prime }]|,$ or 
$(\omega ,\omega ^{\prime },[{\bf k],[k}^{\prime }]|),$ and therefore we
have decoherence in all the dynamical variables.

\subsection{The limit $\hbar \rightarrow 0$ and the classical $\rho
_{*}^{(cl)}(x,k).$}

Let us restore the notation $\{l\}=(\omega ,[{\bf l}]),$ $\{k\}=(\omega ,[%
{\bf k]),}$ as in eq. (\ref{12.18}) and let us consider the configuration
kets $|\{x\}\rangle =|\eta ,[{\bf x}]\rangle .$ Since we are considering the
period when $a\sim a_{out}$ the system with hamiltonian (\ref{12.12}) is
just a set of infinite oscillators with constants $\Omega _{{\bf k}%
}(a_{out}) $ that represent a scalar field with mass $ma_{out}$. Then we are
just dealing with a classical set of $N$ particles, with coordinates $[{\bf x%
}]$ and momenta $[{\bf k}].$ Then, as in eq. (\ref{4.3}) we can introduce
the Wigner function corresponding to generalized state $|\{l\}).$%
\begin{equation}
\rho _{\{l\}}^{(W)}([{\bf x}],[{\bf k}])=\pi ^{-4N}\int (\{l\}|{\bf %
x+\lambda \rangle \langle x-\lambda }|)e^{2i[{\bf \lambda ]\bullet [k]}%
}d^{4n}\lambda  \label{15.1}
\end{equation}
Using the same reasoning that we have used to obtain eq. (\ref{4.6}) it
reads: 
\begin{equation}
\rho _{\{l\}}^{(W)}([{\bf x}],[{\bf k}])=\prod_i\delta (L_i^W([{\bf x}],[%
{\bf k}])-l_i)  \label{15.2}
\end{equation}
where $L_i^W([{\bf x}],[{\bf k}])$ is the classical observable obtained from 
$L_i$ via the Wigner integral (considering $h=L_0$ and including $0$ among
the indices $i).$ Now, with the new notation (\ref{15.1}), eq. (\ref{14.9})
reads: 
\begin{equation}
\rho _{*}=\int \rho _{*}\{l\}(\{l\}|d\{l\}  \label{15.3}
\end{equation}
the if we call: 
\begin{equation}
\rho _{*}^{(W)}([{\bf x}],[{\bf k}])=\pi ^{-4N}\int (\{\rho _{*}|{\bf %
x+\lambda \rangle \langle x-\lambda }|)e^{2i[{\bf \lambda ]\bullet [k]}%
}d^{4n}\lambda  \label{15.4}
\end{equation}
we obtain, as in eq. (\ref{4.1}): 
\begin{equation}
\rho _{*}^{(W)}([{\bf x}],[{\bf k}])=\rho _{*}^{(W)}(L_0^W([{\bf x}],[{\bf k}%
]),L_1^W([{\bf x}],[{\bf k}]),...)  \label{15.5}
\end{equation}
So finally: 
\begin{equation}
\rho _{*}^{(W)}([{\bf x}],[{\bf k}])\sim \int d\{l\}\rho _{\{l\}}\rho
_{*}^{(W)}([{\bf x}],[{\bf k}])\delta (\{L^W\}-\{l\})=\int d\{l\}\rho
_{\{l\}}|\prod_i\delta (L_i^W-l_i)  \label{15.7}
\end{equation}
The last equation can be interpreted as follows:

i.- $\delta (\{p\}-\{l\})$ is a classical density function, strongly peaked
at certain values of the constants of motion $\{l\},$ corresponding to a set
of trajectories, where the momenta are equal to the eigenvalues of eqs. (\ref
{14.9}) and (\ref{14.10}), namely $L_i^W=l_i$ $(i=0,1,2,...)$.

ii.- $\rho _{\{l\}}$ is the probability to be in one of these sets of
trajectories labelled by $\{l\}$. Precisely: if some initial density matrix
is given, from eq. (\ref{14.5}) it is evident that its diagonal terms $\rho
_{\{l\}}$ are the probabilities to be in the states $(\omega ,[{\bf l]|}$
and therefore the probability to find, in the corresponding classical
equilibrium density function $\rho _{*}^{(W)}([{\bf x}],[{\bf k}])$, the
density function $\delta (\{L^W\}-\{l\}),$ namely the probability of the set
of trajectories labelled by $\{l\}=(\omega ,[{\bf l])}.$

iii.- As in eq. (\ref{4.9}) let ${\bf a}$ be the coordinate classically
conjugated to ${\bf l}$ and let be ${\bf a}_0$ the coordinate ${\bf a}$ at
time $\eta =0$, then we obtain the classical trajectories: 
\begin{equation}
{\bf a=l}\eta {\bf +a}_0  \label{15.7'}
\end{equation}

iv.- Let us now call $\rho _{*}\{l\}=p_{\{l\}[{\bf a}_0]}.$ Really $p_{\{l\}[%
{\bf a}_0]}$ is not a function of ${\bf a}_0$, it simply is a constant in $%
{\bf a}_0$, since ${\bf a}_0$ is only an arbitrary point and our model is
spatially homogenous. Then we can write: 
\begin{equation}
p_{\{l\}[{\bf a}_0]}=\int p_{\{l\}[{\bf a}_0]}\prod_{i=1}\delta
(a_i-a_{0i})d[{\bf a}_0]  \label{15.8'}
\end{equation}
in this way we have changed the role of ${\bf a}_0$, it was a fixed (but
arbitrary) point and it is now a variable that moves all over the space.
Then eq. (\ref{15.7}) reads: 
\begin{equation}
\rho _{*}^{(cl)}([{\bf x}],[{\bf k}])\sim \int p_{\{l\}[{\bf a}_0{\bf ]}%
}\prod_i\delta (L_i^W-l_i)\prod_{j=1}\delta (a_j-a_{0j})d[{\bf a}_0]d\{l\}
\label{15.8'''}
\end{equation}
So if we call : 
\begin{equation}
\rho _{\{l\}[{\bf a}_0]}^{(cl)}([{\bf x}],[{\bf k}])=\prod_{i=0}\delta
(L_i^W-l_i)\prod_{j=1}\delta (a_j-a_{0j})  \label{15.8''''}
\end{equation}
we have: 
\begin{equation}
\rho _{*}^{(cl)}([{\bf x}],[{\bf k}])\sim \int p_{\{l\}[{\bf a}_0{\bf ]}%
}\rho _{\{l\}[{\bf a}_0]}^{(cl)}([{\bf x}],[{\bf k}])d[{\bf a}_0{\bf ]}d\{l\}
\label{15.8'''''}
\end{equation}
From eq. (\ref{15.8''''}) we see that $\rho _{\{l\}[{\bf a}_0]}^{(cl)}([{\bf %
x}],[{\bf k}])\neq 0$ only in a narrow strip around the classical trajectory
(\ref{15.7'}) defined by the momenta $\{l\}$ and passing through the point [$%
{\bf a}_0{\bf ]}$ (really the density function is as peaked as it is allowed
by the uncertainty principle, so its width is essentially a $O(\hbar ),$
since the $\delta -$functions of all the equation are really Dirac's deltas
only when $\hbar \rightarrow 0)$ . So we have proved eq. (\ref{15.8'''''})
which, in fact, it is eq. (\ref{1.3}) as announced \footnote{%
In this section, as in sectionn IIB, we have faced the following problem:
\par
$\rho _{*}^{(cl)}([{\bf x}],[{\bf k}])$ is a ${\bf a}$ constant that we want
to decompose in functions $\rho _{\{l\}[{\bf a}_0]}^{(cl)}([{\bf x}],[{\bf k}%
])$ which are different from zero only around the trajectory (\ref{15.7'})
and therefore are variables in ${\bf a.}$%
\par
Then,essentially we use the fact that if $f(x,y)=g(y)$ is a constant
function in $x$ we can decompose it as: 
\[
g(y)=\int g(y)\delta (x-x_0)dx_0 
\]
namely the densities $\delta (x-x_0)$ are peaked in the trajectories $%
x=x_0=const.,y=var.$ and, therefore, are functions of $x.$ This trajectories
play the role of those of eq. (\ref{15.8'}).
\par
As all the physics, including the correlations, is already contained in eq. (%
\ref{15.7}), the reader may just consider the final part of this section,
from eq. (\ref{15.8'}) to eq. (\ref{15.8'''''}) a didactical trick.}.

Then we have obtained the classical limit. When $\eta \rightarrow \infty $
the quantum density $\rho $ becomes a diagonal density matrix $\rho _{*}.$
The corresponding classical distribution $\rho _{*}^{(cl)}([{\bf x}],[{\bf k}%
])$ can be expanded as a sum of classical trajectories density functions $%
\rho _{\{l\}[{\bf a}_0]}^{(cl)}([{\bf x}],[{\bf k}]),$ each one weighted by
its corresponding probability $p_{\{l\}[{\bf a}_0{\bf ]}}.$ So, as the limit
of our quantum model we have obtained a statistical classical mechanical
model, and the classical realm appears.

\subsection{Localization and correlations.}

Under adequate initial conditions the motion can be concentrated in just one
trajectory showing, the presence of correlations in this trajectory \cite
{LauraF}. The evolution of the concentration depends on potential $V(a).$ Of
course, our ''trajectories'' are not only one trajectory for a one particle
state, but they are $N$ trajectories (each one corresponding to a momenta $%
(l_1,l_2,...l_n)=\{l\}$ and passing by a point (${\bf a}_1,{\bf a}_2,...,%
{\bf a}_n)=[{\bf a])}$ for the n particle states.

\subsection{Discussion and comments.}

\subsubsection{Characteristic times.}

The decaying term of eq. (\ref{2.8}) (i. e. the second term of the r. h. s.)
can be analytically continued using the techniques explained in papers \cite
{LauraA}, \cite{CyLI}, and \cite{Cast}. In these papers it is shown that
each pole $z_i=\omega _i-i\gamma _i,$ of the S-matrix (corresponding to the
evolution $a_{in}\rightarrow a_{out}$, see \cite{Cast}$)$ of the problem
considered, originates a damping factor $e^{-\gamma _i\eta }.$ Then if $%
\gamma =\min (\gamma _i)$ the characteristic decoherence time is $\gamma
^{-1}.$ This computation is done in the specific models of papers \cite{Cast}%
. If $\gamma \ll 1,$ even if the Riemann-Lebesgue theorem is always valid,
there is no practical decoherence since $\gamma ^{-1}\gg 1.$

\subsubsection{Sets of trajectories decoherence.}

It is usual to say that in the classical regime there is decoherence of the
set trajectories labelled by the constant of the motion $\omega ,$ $[{\bf l}%
] $. This result can easily be obtained with our method in the following way.

i.- Let us consider two different states $|\omega [{\bf l}]\rangle $ and $%
|\omega ^{\prime }[{\bf l}^{\prime }]\rangle $ that will define classes of
trajectories with different constants of the motion $(\omega ,[{\bf l}])\neq
(\omega ^{\prime },[{\bf l}^{\prime }]).$ We must compute: 
\begin{equation}
\langle \omega [{\bf l}]|\rho _{*}|\omega ^{\prime }[{\bf l}^{\prime
}]\rangle =(\rho _{*}||\omega \omega ^{\prime }[{\bf l}][{\bf l}^{\prime
}])=\left[ \int \rho _{\omega ^{\prime \prime }[{\bf l}^{\prime \prime
}]}(\omega ^{\prime \prime }[{\bf l}^{\prime \prime }]|d\omega ^{\prime
\prime }d[{\bf l}^{\prime \prime }]\right] |\omega \omega ^{\prime }[{\bf l}%
][{\bf l}^{\prime }])=0  \label{16.1}
\end{equation}
due to the orthogonality of the basis $\{(\omega ,[{\bf l]}|,(\omega ,\omega
^{\prime },[{\bf l]|}\}$ .

ii.- But if we compute: 
\[
\langle \omega [{\bf l}]|\rho _{*}|\omega [{\bf l}]\rangle =(\rho
_{*}||\omega [{\bf l}])=\left[ \int \rho _{\omega ^{\prime \prime }[{\bf l}%
^{\prime \prime }]}(\omega ^{\prime \prime }[{\bf l}^{\prime \prime
}]|d\omega ^{\prime \prime }d[{\bf l}^{\prime \prime }]\right] |\omega [{\bf %
l}])= 
\]
\begin{equation}
\int \rho _{\omega ^{\prime \prime }[{\bf l}^{\prime \prime }]}\delta
(\omega -\omega ^{\prime \prime })\delta ([{\bf l}]-[{\bf l}^{\prime \prime
}])d\omega ^{\prime \prime }d[{\bf l}^{\prime \prime }]=\rho _{\omega [{\bf l%
}]}\neq 0  \label{16.2}
\end{equation}
The last two equations complete the demonstration. We will discuss the
problem of the decoherence of two trajectories, with the same $\{l\}$ but
different $[{\bf a}_0]$ in subsection IVD4.

\subsubsection{A discussion on time decoherence.}

It is well known that one of the main problems of quantum gravity is the
problem of the time definition (see \cite{Time}). A not well studied feature
of this problem is that, there must be a decoherence process related with
time, since time is as a classical variable. In this subsection, using the
functional technique, we will give a model that shows that this is the case
(but we must emphasize that this subject is not completely developed).

Let us postulate that there is a parameter $\eta $ such that the quantum
states evolve as \footnote{%
Of course $\eta $ is the conformal time of eq. (\ref{12.10}), since (\ref
{16.3}) is a consequence of (\ref{12.10}). But now we have postulated this
last equation and we are searching the quantum properties of $\eta .$}:

\begin{equation}
|\eta \rangle =e^{-ih\eta }|0\rangle  \label{16.3}
\end{equation}

We must compute $\langle \eta |\rho _{*}|\eta ^{\prime }\rangle $ where $%
|\eta \rangle $ and $|\eta ^{\prime }\rangle $ are two states of the system
for different times. We do not know if $\langle \eta |\rho _{*}|\eta
^{\prime }\rangle $ will decohere or not. If it decohers we can say that
parameter $\eta $ is a classical one. $|\eta \rangle \langle \eta ^{\prime
}| $ can be considered as an observable, then: 
\begin{equation}
\langle \eta ^{\prime }|\rho _{*}|\eta \rangle =(\rho _{*}||\eta \rangle
\langle \eta ^{\prime }|)  \label{16.4}
\end{equation}
But: 
\begin{equation}
(\omega ||\eta \rangle \langle \eta ^{\prime }|)=(\omega |e^{-ih\eta
}|0\rangle \langle 0|e^{ih\eta ^{\prime }})=[e^{ih\eta ^{\prime }}(\omega
|e^{-ih\eta }]||0\rangle \langle 0|)  \label{16.5}
\end{equation}
Now, for any observable $O$ we have: 
\[
\lbrack e^{ih\eta ^{\prime }}(\omega |e^{-ih\eta }]||O)=[e^{ih\eta ^{\prime
}}(\omega |e^{-ih\eta }]|[\int O_{\omega ^{\prime }}|\omega ^{\prime
})d\omega ^{\prime }+\int \int O_{\omega ^{\prime }\omega ^{\prime \prime
}}|\omega ^{\prime },\omega ^{\prime \prime })d\omega ^{\prime }d\omega
^{\prime \prime })= 
\]
\[
\lbrack e^{ih\eta ^{\prime }}(\omega |e^{-ih\eta }]|[\int O_{\omega ^{\prime
}}|\omega ^{\prime })d\omega ^{\prime }+...=(\omega |[\int O_{\omega
^{\prime }}e^{-i\omega ^{\prime }\eta }|\omega ^{\prime })e^{i\omega
^{\prime }\eta ^{\prime }}d\omega ^{\prime }])= 
\]
\begin{equation}
e^{-i\omega (\eta ^{\prime }-\eta )}(\omega |O)  \label{16.6}
\end{equation}
where the second term disappears since $(\omega |\omega ^{\prime },\omega
^{\prime \prime })=0.$ Thus: 
\begin{equation}
(\omega ||\eta \rangle \langle \eta ^{\prime }|)=e^{-i\omega (\eta ^{\prime
}-\eta )}(\omega ||0\rangle \langle 0|)  \label{16.7}
\end{equation}
So now we can compute the following two cases:

i.- 
\begin{equation}
\langle \eta ^{\prime }|\rho _{*}|\eta \rangle =(\rho _{*}||\eta \rangle
\langle \eta ^{\prime }|)=[\int \rho _\omega (\omega |d\omega ]||\eta
\rangle \langle \eta ^{\prime }|)=\int \rho _\omega e^{-i\omega (\eta
^{\prime }-\eta )}(\omega ||0\rangle \langle 0|)d\omega \rightarrow 0
\label{16.8}
\end{equation}
when%
\mbox{$\vert$}%
$\eta ^{\prime }-\eta |\rightarrow \infty $ , due to the Riemann-Lebesgue
theorem$.$

ii.- Analogously:

\begin{equation}
\langle \eta |\rho _{*}|\eta \rangle =\int \rho _\omega (\omega ||0\rangle
\langle 0|)d\omega \neq 0  \label{16.9}
\end{equation}
So we have time decoherence for two times $\eta $ and $\eta ^{\prime }$ if
they are far enough.

This result is important for the problem of time definition, since in order
to have a reasonable classical time this variable must first decohere. The
result above shows that this is the case for $\eta $ and $\eta ^{\prime }$
far enough \footnote{%
Using the method of section IVC1 we can compute $\gamma .$ Decoherence will
take place for $|\eta -\eta ^{\prime }|>\gamma ^{-1}.$}, but also that, for
closer times (namely such that their difference is smaller than Planck's
time) there is no decoherence and time cannot be considered as a classical
variable. Classical time is a familiar concept but the real nature of the
non-decohered quantum time is opened to discussion. But we must remark that,
some how, we have followed the second line of thought of section IIC1: we
have suppose the existence of an evolution $e^{-ih\eta },$ where $\eta $ is
only a parameter. We have prove that decoherence appears and find that $\eta 
$ behaves like a classical variable. May be this is the better way to
introduce the classical time to postulate a ''quantum'' $\eta $ and to find
its properties. Moreover we have proved that the second line of section IIC1
can as well be followed.

\subsubsection{Decoherence in the space variables.}

Now that we know that there is time decoherence we can repeat the reasoning
for the rest of the variables ${\bf a}$ at time $\eta =0$ and changing eq. (%
\ref{16.3}) by: 
\begin{equation}
|[{\bf a]\rangle =}e^{i[{\bf a]\bullet [l]}}|{\bf 0\rangle }  \label{16.10}
\end{equation}
and we will reach to the conclusions:

i.- 
\begin{equation}
\langle [{\bf a]}|\rho _{*}|[{\bf a}^{\prime }]\rangle \rightarrow 0
\label{16.11}
\end{equation}
when 
\mbox{$\vert$}%
${\bf a-a}^{\prime }|\rightarrow \infty .$

ii,- 
\begin{equation}
\langle [{\bf a]}|\rho _{*}|[{\bf a]}\rangle \neq 0  \label{16.12}
\end{equation}
therefore there is also decoherence between two trajectories with the same $%
\{l\}$ but different $[{\bf a}_0]$.

These facts complete the scenario about decoherence and the final classical
limit. An analogy of section IIIC2 in the cosmological case can be found in
paper \cite{Castagnino}.

\section{Conclusion.}

We are convinced that the method of papers \cite{LauraA} and \cite{LauraE}
is the best way to study both the Mott and the cosmological problem and to
find the analogies and differences between then. We hope that the reader
will share our conviction.

For the most important of both model, the cosmological one we have shown
that after the WKB expansion and the decoherence and the final classical
limit process our quantum model has:

i.-A defined classical time $\eta $ and a defined classical geometry related
by eq. (\ref{12.10}).

ii.- Decoherence has appeared in a well defined final pointer basis.

iii.- The quantum field has originated a classical final distribution
function (eq. (\ref{15.8'''''})) that is a weighted average of some set
densities, each one related to a classical trajectory. The weight
coefficients are the probabilities of each trajectory.

We can foresee that if instead of a spinless field we would coupled the
geometry with a spin 2 metric fluctuation field the result would be more or
less the same. Then the corresponding quantum fluctuations would become
classical fluctuations that would correspond to matter inhomogeneities
(galaxies, clusters of galaxies, etc.) that will move along the trajectories
described above. But this subject will be treated elsewhere with greater
detail.

\section{Acknowledgments.}

We are very grateful to Julien Barbour who points us the relation between
the Mott and the cosmological problems. This work was partially supported by
grants CI1$^{*}$-CT94-0004 and PSS$^{*}-0992$ of the European Community, PID
3183/93 of CONICET, EX053 of the Buenos Aires University, and also grants
from Fundaci\'{o}n Antorchas and OLAM Foundation.

\end{document}